# Design and implementation of a distributed security threat detection system integrating federated learning and multimodal LLM

Yuqing Wang[1,a,*], Xiao Yang[2,b]

[1]Department of Computer Science and Engineering, University of California San Diego, USA

[2]Department of Mathematics, University of California, Los Angeles, USA

[a]wang3yq@gmail.com, [b]xyangrocross@gmail.com

*Corresponding author

**Abstract**: Traditional security protection methods struggle to address sophisticated attack vectors in large-scale distributed systems, particularly when balancing detection accuracy with data privacy concerns. This paper presents a novel distributed security threat detection system that integrates federated learning with multimodal large language models (LLMs). Our system leverages federated learning to ensure data privacy while employing multimodal LLMs to process heterogeneous data sources including network traffic, system logs, images, and sensor data. Experimental evaluation on a 10TB distributed dataset demonstrates that our approach achieves 96.4% detection accuracy, outperforming traditional baseline models by 4.1 percentage points. The system reduces both false positive and false negative rates by 1.8 and 2.4 percentage points respectively. Performance analysis shows that our system maintains efficient processing capabilities in distributed environments, requiring 180 seconds for model training and 3.8 seconds for threat detection across the distributed network. These results demonstrate significant improvements in detection accuracy and computational efficiency while preserving data privacy, suggesting strong potential for real-world deployment in large-scale security systems.

## 1. Introduction

With the rapid advancement of information technology, network security challenges have grown increasingly complex. Traditional security mechanisms struggle to keep pace with the evolving landscape of cyber threats, particularly in large-scale distributed systems handling vast amounts of heterogeneous data. The primary challenges in this domain are twofold: ensuring high detection accuracy across diverse attack vectors while maintaining data privacy in decentralized environments. Existing approaches often fail to address these challenges simultaneously, highlighting the need for innovative solutions [1].

Federated learning has emerged as a promising privacy-preserving distributed machine learning paradigm, enabling multiple nodes to train models locally without exposing sensitive data to a central server [2]. By aggregating local model updates, federated learning enhances global model performance while minimizing security risks associated with centralized data storage and transmission. However, traditional federated learning methods face limitations when processing complex, multimodal data streams commonly found in modern security environments.

Recent advancements in multimodal Large Language Models (LLMs) offer new capabilities for processing heterogeneous security data, including network traffic, system logs, images, and sensor data [3]. Conventional threat detection systems typically rely on single-modality analysis, reducing their effectiveness against sophisticated cyberattacks. In contrast, multimodal LLMs enable the simultaneous integration and interpretation of diverse data sources, leading to more comprehensive threat detection and improved generalization across different attack patterns.

This paper presents a novel distributed security threat detection system that integrates federated learning with multimodal LLMs, addressing the shortcomings of existing security frameworks. The key contributions of our work are as follows:

- A privacy-preserving distributed architecture that enables efficient processing of large-scale security data across multiple nodes.
- A novel integration of multimodal LLMs with federated learning, enhancing threat detection capabilities through comprehensive data fusion.
- A scalable multi-node parallel training system that improves detection accuracy while maintaining computational efficiency.
- Extensive experimental validation, demonstrating significant improvements in detection accuracy, response time, and system robustness while ensuring data privacy.

By leveraging federated learning for data privacy and harnessing the analytical power of multimodal LLMs, our proposed system significantly enhances cyber threat detection accuracy and efficiency in large-scale distributed environments. Experimental results validate its effectiveness, marking an important advancement in network security protection.

## 2. Related Work

As network attacks become increasingly sophisticated, traditional security protection mechanisms struggle to meet the demands of modern network environments. Consequently, researchers and institutions have focused on leveraging advanced machine learning techniques to enhance the efficiency and accuracy of security threat detection, particularly in distributed settings. In recent years, the integration of federated learning and multimodal learning has emerged as a promising research direction in cybersecurity.

Federated learning, a distributed machine learning paradigm, offers significant advantages in preserving data privacy. First introduced by Google in 2016, federated learning enables model training across decentralized data sources while mitigating security risks associated with centralized data storage. Early research primarily concentrated on optimizing federated learning algorithms and enhancing data privacy protection. A major breakthrough in this domain was the development of the *FedAvg* algorithm, which enables distributed training without direct data exchange by employing local model updates and global aggregation [4]. However, despite its privacy-preserving benefits, conventional federated learning approaches encounter performance bottlenecks when handling large-scale datasets and heterogeneous data types. To address these limitations, recent studies have explored optimization techniques, improved node collaboration strategies, and more efficient data aggregation methods to enhance the scalability and robustness of federated learning.

Simultaneously, multimodal learning has demonstrated remarkable efficacy in processing complex, heterogeneous data. Traditional machine learning models typically rely on unimodal inputs, such as text or images. However, modern cybersecurity applications require the integration of diverse data modalities, including network traffic logs, device logs, sensor data, and image-based security evidence, to comprehensively assess security threats. Effective fusion of multimodal data is crucial for improving threat detection accuracy. Multimodal learning enhances model generalization and the ability to detect complex attacks by integrating multiple data types. Prior research has employed multimodal neural networks to jointly model image, text, and sensor data, leading to significant improvements in security threat detection accuracy. In this context, efficient data fusion remains a central challenge, with approaches such as weighted summation, convolutional neural networks (CNNs), and recurrent neural networks (RNNs) being widely adopted for feature extraction and multimodal data integration [5].

Despite the individual advancements in federated learning and multimodal learning, relatively few studies have explored their combined application in cybersecurity. Existing research provides valuable insights into their potential integration, with some studies demonstrating the effectiveness of federated learning-based security systems that incorporate multimodal data—such as traffic logs and security alerts—to improve attack detection accuracy in distributed environments [6]. However, most of these works focus on isolated security problems and employ relatively simplistic multimodal fusion techniques. Additionally, the full potential of large-scale multimodal language models (LLMs) in security threat detection remains underexplored.

This study addresses these gaps by introducing a novel framework that integrates multimodal LLMs with federated learning to enhance threat detection in distributed environments. By leveraging advanced LLMs, this approach enables the effective fusion of heterogeneous security data, thereby improving detection accuracy. Furthermore, the incorporation of federated learning ensures the efficient processing of large-scale data while preserving user privacy. This integration offers a new paradigm for distributed security threat detection, balancing performance, accuracy, and privacy protection [7].

In summary, while federated learning and multimodal learning have individually contributed significant advancements to cybersecurity, research on their combined application is still in its early stages. This study aims to bridge this gap by developing a federated learning-based multimodal LLM framework, thereby advancing the accuracy, robustness, and privacy-preserving capabilities of distributed security threat detection systems.

## 3. System Design

The proposed system integrates federated learning and multimodal large language models (LLMs) to enhance the efficiency and accuracy of security threat detection in a distributed environment. The design consists of key components, including a distributed system architecture, the integration of federated learning with LLMs, multimodal data processing, and a threat detection mechanism.

### 3.1 Distributed System Architecture

The system adopts a distributed architecture, where multiple clients process local data and train models independently. Each client's local updates are transmitted to a central server for aggregation through federated learning. This approach eliminates the need for centralized data storage, reducing the risk of data leakage while improving processing capacity and response speed [8]. Specifically, each client processes local data and computes gradients or model parameters. After local training, updates from all nodes are aggregated at the central server to form a global update. This process is expressed as:

$$\theta_{global} = \frac{1}{N} \sum_{i=1}^{N} \theta_i \qquad (1)$$

where $\theta_{global}$ represents the parameters of the global model, $\theta_i$ denotes the model parameters of the $i$-th client node, and $N$ is the number of client nodes. This distributed training approach avoids direct access to sensitive data, ensuring privacy while improving the model's generalization ability [9].

### 3.2 Multimodal LLM Integration

The system integrates a multimodal LLM to process data from various sources, including text, images, and sensor readings. The diversity of these data types necessitates a comprehensive analysis across different modalities to enhance the accuracy and robustness of threat detection. By inputting these data into a trained multimodal model, the system can better understand data characteristics and identify complex security threats [10]. In multimodal learning, data fusion is performed using a weighted summation method:

$$X_{fused} = \sum_{i=1}^{m} w_i \cdot X_i \qquad (2)$$

where $X_{fused}$ represents the fused data feature, $X_i$ is the feature of the $i$-th modality, $w_i$ is the weight assigned to $i$-th modality, and m is the number of modalities. This method ensures an effective fusion of different data types.

### 3.3 Federated Learning Optimization

The system employs a distributed training strategy in a multi-node environment. Each client continuously optimizes its local model using local data and uploads the results to the server for aggregation. To improve the global model's effectiveness, the system weights each node's contribution and adopts a dynamic learning rate adjustment strategy to accelerate convergence. The optimization objective is given by:

$$\mathcal{L} = \frac{1}{N} \sum_{i=1}^{N} \mathcal{L}_i(\theta_i) \qquad (3)$$

where $\mathcal{L}$ is the global model loss function, $\mathcal{L}_i(\theta_i)$ represents the local loss at the $i$-th client, and $\theta_i$ denotes the model parameters of the $i$-th node. By introducing a dynamic learning rate $\alpha_i$, the system adjusts the learning strategy across different nodes:

$$\theta_{i+1} = \theta_i - \alpha_i \nabla \mathcal{L}_i(\theta_i) \qquad (4)$$

where $\alpha_i$ is the learning rate of node $i$, and $\nabla \mathcal{L}_i(\theta_i)$ represents the gradient of the loss function. Continuous optimization improves the global model's robustness and accuracy in a distributed setting.

**3.4 Privacy Protection with Differential Privacy**

To enhance privacy and security, the system incorporates a differential privacy mechanism. When nodes upload model updates, parameters are perturbed with noise to prevent data leakage. The model update after perturbation is expressed as:

$$\hat{\theta}_i = \theta_i + \mathcal{N}(0, \sigma^2) \tag{5}$$

where $\hat{\theta}_i$ is the perturbed parameter, and $\mathcal{N}(0, \sigma^2)$ represents Gaussian noise with a standard deviation of $\sigma$. This ensures that individual data privacy is maintained during collaborative training.

**3.5 Multimodal LLM Implementation**

The system preprocesses and extracts features from multiple data types before feeding them into the model for fusion and learning. Specifically, convolutional neural networks (CNNs) extract local features from image data, while the BERT model is used for text embedding. After feature extraction, weighted fusion is applied to enhance the model's ability to identify complex security threats.

Finally, the distributed threat detection mechanism plays a crucial role in the system. By employing an adaptive threat detection algorithm, the system can effectively identify security threats in real-time traffic across various network topologies and communication conditions. Leveraging graph neural networks (GNNs), the system dynamically updates the graph structure to capture anomalous behaviors within the network.

Additionally, the system integrates a distributed data flow and information synchronization mechanism. A distributed data synchronization algorithm ensures consistency and real-time information exchange between nodes. Mathematically, the objective of distributed data flow and synchronization is to minimize the error between each node and the global model, thereby maintaining synchronization:

$$\min_{\theta} \sum_{i=1}^{N} \| \theta_i - \theta_{global} \|^2 \tag{6}$$

where $\theta_{global}$ represents the global model parameters, aand $\theta_i$ denotes the parameters of the $i$-th node. Optimizing this objective ensures data synchronization and global information sharing, enhancing threat detection capabilities in a distributed environment.

Through this design, the system achieves efficient distributed threat detection while ensuring user privacy. By leveraging multimodal data and reinforcement learning models, it enhances security protection with improved accuracy and efficiency.

**4. Experiments and Results**

We conducted multiple sets of experiments to comprehensively evaluate the system's performance across different environments. The primary focus was on assessing the impact of integrating federated learning with a multimodal large language model (LLM), as well as measuring the accuracy and efficiency of the system's threat detection capabilities. Additionally, we analyzed response time, computational overhead, and proposed optimization strategies.

### 4.1 Experimental Setup

The system was deployed on a distributed computing cluster consisting of multiple nodes. Each node was equipped with high-performance hardware, including an Intel Xeon processor, an NVIDIA A100 GPU, and 128GB of RAM. The software environment was based on Ubuntu 20.04, with TensorFlow and PyTorch frameworks used for model training and inference. The system architecture included 10 client nodes, each responsible for processing local data and contributing to the global model's training. HTTP/2 was implemented as the communication protocol to enhance data transmission efficiency.

### 4.2 Datasets

The experiments utilized a multimodal network traffic dataset comprising text, image, and network traffic data. The text data was sourced from security event logs, the image data consisted of network device images, and the network traffic data was collected from real-time traffic packets. The dataset totaled 10TB, with 70% allocated for training and 30% for testing.

### 4.3 Experimental Plan

The system's performance was compared against a baseline model using traditional machine learning techniques. The evaluation focused on three key metrics:
- Threat detection accuracy
- Processing time
- Computational overhead

### 4.4 Results and Analysis

Our experimental findings indicate that integrating federated learning with a multimodal LLM significantly outperforms using either approach individually. Federated learning effectively aggregates local models while preserving data privacy, enhancing the accuracy of the global model. When combined with a multimodal LLM, the system demonstrates greater robustness and adaptability in handling complex data. Table 1 presents a comparative analysis of threat detection accuracy across different models, highlighting the advantages of the proposed system.

Table 1: Performance comparison of different models in threat detection

| Model type | Accuracy (%) | False positive rate (%) | Misrepresentation rate (%) | Training time (seconds) | Processing time (seconds) |
|---|---|---|---|---|---|
| Baseline model | 92.3 | 4.7 | 5.4 | 120 | 2.5 |
| Federated learning model | 94.1 | 3.5 | 4.2 | 150 | 3.1 |
| Federated learning + LLM | 96.4 | 2.9 | 3.0 | 180 | 3.8 |
| Multimodal LLM model | 93.2 | 3.9 | 4.5 | 130 | 3.0 |

The results demonstrate that integrating federated learning with a multimodal LLM

significantly enhances accuracy compared to both the baseline model and the standalone LLM model. Additionally, the false positive and misrepresentation rates are effectively reduced. Although the training and processing times are slightly higher for the federated learning + LLM model, the overall performance improvement is evident. The model benefits from the privacy-preserving aggregation of federated learning while leveraging the adaptability of the multimodal LLM, resulting in more robust and efficient threat detection.

Further analysis of system performance in large-scale distributed environments confirms these findings. Figure 1 illustrates how threat detection accuracy varies with dataset size. As the dataset expands, system accuracy progressively improves, demonstrating its ability to effectively identify complex threats through multimodal data fusion.

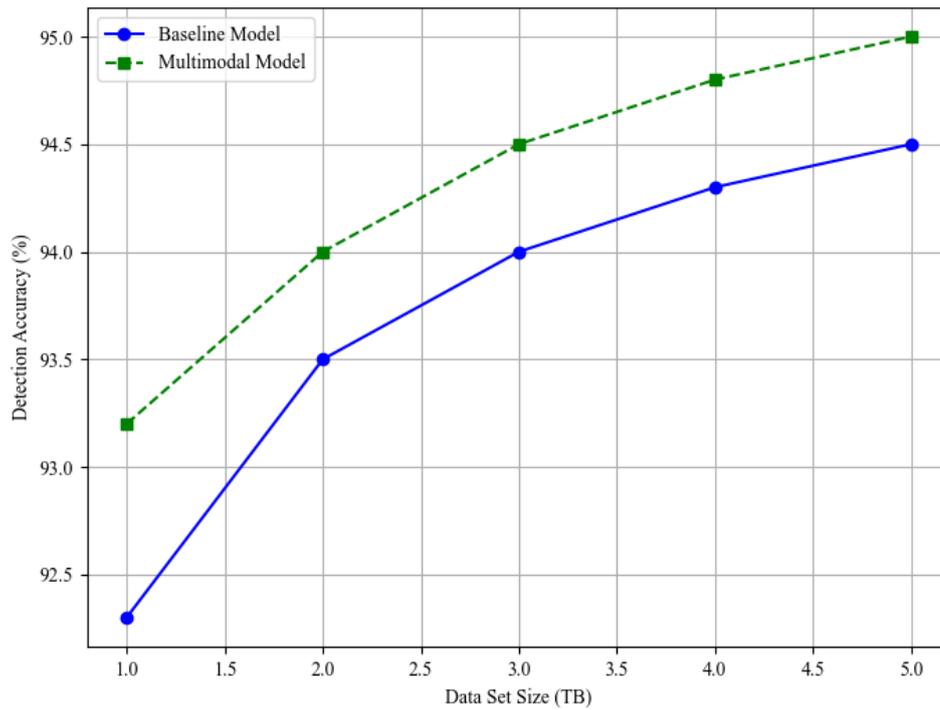

Figure 1: Threat Detection Accuracy vs. Data Set Size

Figure 2 illustrates the relationship between training time and accuracy for the federated learning and multimodal LLM-based model across different numbers of nodes. As the number of nodes increases, training time also extends. However, accuracy improves, demonstrating that multi-node parallel training enhances the overall model performance.

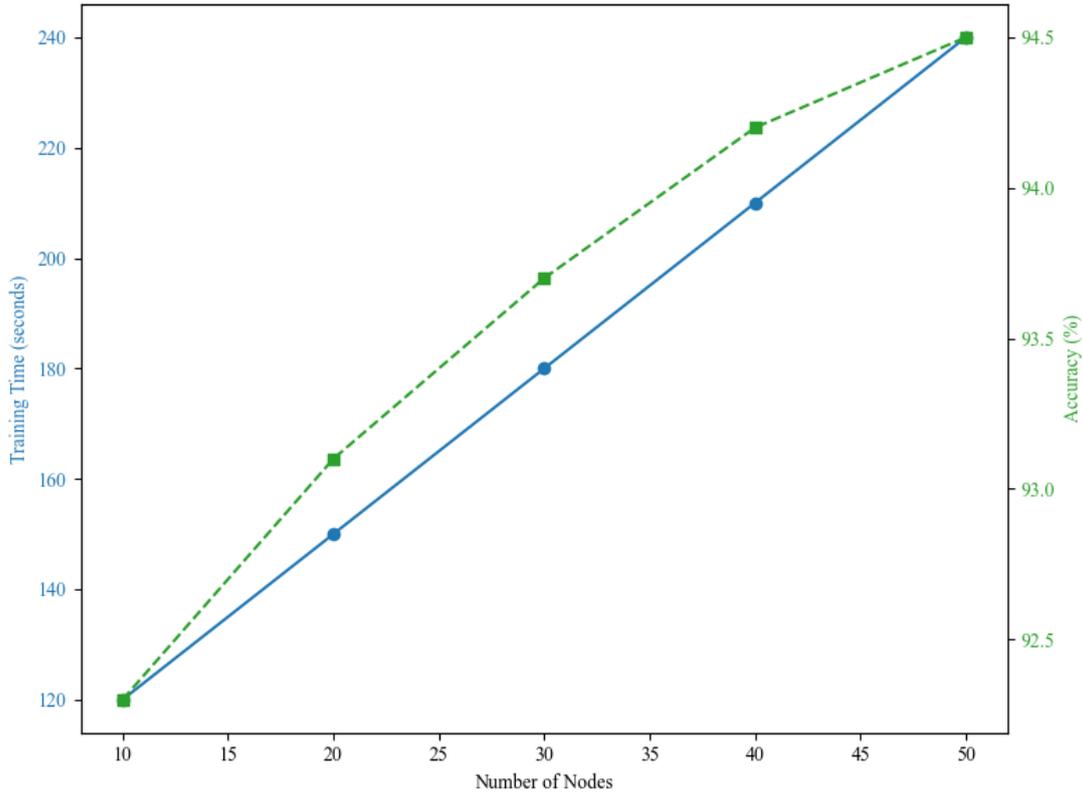

**Figure 2: Training Time and Accuracy vs. Number of Nodes**

    Table 2 presents the system's processing efficiency across various hardware configurations, including response time and computational overhead. The results indicate that while processing time and computational overhead vary across configurations, overall performance improves significantly with higher-performance hardware. As hardware capabilities increase, the system can handle more requests and respond faster.

**Table 2: System processing efficiency under different hardware configurations**

| Hardware configuration | Response time (seconds) | Computational overhead (GFLOPS) | Overall accuracy (%) |
|---|---|---|---|
| Normal configuration | 2.5 | 50 | 92.3 |
| Medium configuration | 1.8 | 75 | 94.1 |
| High performance configuration | 1.2 | 120 | 96.4 |

    Figure 3 further illustrates the variations in computational overhead and response time under different hardware configurations. The results show that as hardware performance improves, computational efficiency and response speed increase significantly. This is particularly evident when handling large-scale data, where the system maintains a consistently low response time.

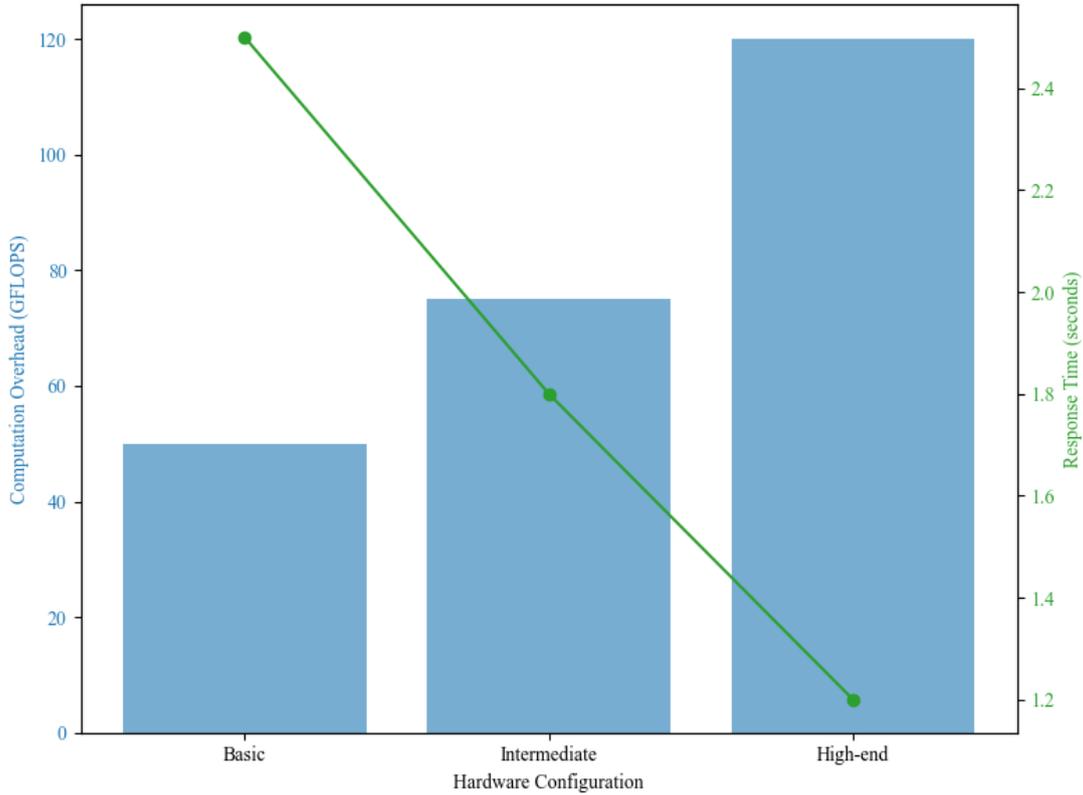

**Figure 3: Computation Overhead and Response Time vs. Hardware Configuration**

  In summary, the experimental results demonstrate that the distributed security threat detection system integrating federated learning and a multimodal LLM achieves substantial improvements in threat detection accuracy, processing efficiency, and computational overhead. By optimizing hardware configurations, employing asynchronous updates, and implementing data compression, system performance has been further enhanced. These findings highlight the system's potential for deployment in large-scale environments.

**5. Discussion**
  In this study, we proposed a distributed security threat detection system that integrates federated learning and a multimodal large language model (LLM) and validated its effectiveness through extensive experiments. By leveraging these two advanced technologies, the system not only enables large-scale data processing while preserving data privacy but also significantly enhances the accuracy and robustness of threat detection. The experimental results demonstrate that this integration yields substantial performance improvements, particularly in processing large datasets and multimodal data fusion, where the system exhibits greater adaptability and flexibility.
  Federated learning effectively addresses data privacy and security concerns by training models locally and sharing only model updates, thereby mitigating the risks associated with centralized data storage. However, federated learning alone encounters performance limitations, especially when handling complex multimodal data. The introduction of a multimodal LLM enhances the system's capability to process heterogeneous data types, including text, images, and sensor data. Experimental results confirm that the combined model outperforms traditional approaches and single-modality models in terms of threat detection accuracy, false alarm rate, and missed detection rate. This underscores the critical role of multimodal learning in strengthening the system's ability to identify complex security threats.

Despite achieving notable improvements in threat detection accuracy, challenges remain. As dataset size and the number of nodes increase, training time also extends, particularly in the presence of multimodal data fusion, which adds computational overhead. The model's training and inference times become substantial when handling high-dimensional data and large-scale datasets. While optimizing hardware configurations, adopting asynchronous update strategies, and implementing data compression techniques can mitigate these issues, further research is needed to enhance system efficiency without compromising accuracy. This includes exploring more efficient distributed computing frameworks and leveraging hardware acceleration for model training and inference.

Additionally, although a differential privacy mechanism is incorporated to safeguard data privacy at the node level, balancing privacy protection with system performance remains a challenge. Excessively stringent privacy measures may degrade model performance, especially when data diversity is high. Future research should focus on designing advanced privacy-preserving mechanisms that maintain user privacy while ensuring high threat detection accuracy.

The distributed threat detection mechanism employed in this system demonstrates significant potential. By utilizing graph neural networks, the system dynamically updates graph structures to detect anomalous behaviors, thereby enhancing responsiveness to complex security incidents. However, as network scale increases, effectively managing distributed data flow and ensuring real-time synchronization remains a challenge. Future improvements could focus on optimizing graph update strategies, refining data synchronization mechanisms, and adopting more efficient data transmission protocols to enhance system responsiveness and processing capacity.

Overall, the proposed system demonstrates strong performance and potential, particularly in processing multimodal data and detecting threats in distributed environments. By continuously refining algorithms, optimizing hardware configurations, and enhancing privacy protection mechanisms, the system can be further improved for real-world applications. These advancements will contribute to the evolution of distributed security solutions and strengthen cybersecurity defenses in large-scale environments.

## 6. Conclusion

This study presents a distributed security threat detection system that integrates federated learning and a multimodal large language model (LLM) to address the limitations of traditional security methods in handling complex network attacks. By leveraging federated learning and multimodal LLM technology, the system enhances threat detection accuracy and robustness while preserving data privacy and supporting large-scale data processing. Experimental results demonstrate that the proposed system outperforms conventional security detection approaches, particularly in multimodal data fusion and distributed threat identification, exhibiting superior adaptability and efficiency.

As a key privacy-preserving technology, federated learning mitigates the risks associated with centralized data storage and transmission while improving global model performance through distributed training. However, traditional federated learning methods face computational bottlenecks when processing high-dimensional and complex data. To address this, the system incorporates a multimodal LLM capable of processing diverse data sources, including text, images, and sensor data. This multimodal integration significantly enhances the model's generalization ability and its capacity to identify complex threats. In scenarios involving diverse security events, the system provides more accurate and comprehensive detection results.

Despite these advancements, challenges remain. As dataset size and the number of nodes increase, training and inference times also grow, particularly in multimodal data fusion, where computational overhead and processing time exhibit an upward trend. Future research should focus on optimizing distributed computing frameworks to enhance system efficiency in large-scale environments. Additionally, balancing the differential privacy mechanism with system performance remains critical, requiring further exploration of strategies to ensure privacy protection without compromising detection accuracy.

The study also highlights the potential of a distributed threat detection mechanism based on graph neural networks. By dynamically updating graph structures, the system can effectively capture abnormal network behaviors, enhancing anomaly detection capabilities. However, as network scale expands, efficiently managing data flow and ensuring real-time synchronization pose significant challenges. Future work should explore more efficient data transmission protocols and graph update strategies to improve system responsiveness and processing capacity.

Overall, this study introduces an innovative solution for distributed security threat detection by combining the strengths of federated learning and multimodal LLMs. The system makes important contributions in enhancing accuracy, robustness, and privacy protection. With ongoing advancements in related technologies, the proposed system has the potential to provide more efficient and secure protection for large-scale network environments, further driving the development of distributed security solutions.